\documentclass[a4paper,12pt, epsfig]{article}
\def\letter{0}\def\pr{0}
\pdfoutput=1 
\usepackage{epsfig}
\usepackage{epstopdf}
\usepackage{textcomp}
\usepackage{graphicx}
\usepackage{ifthen}
\usepackage{ulem}

\usepackage{amsmath}
\usepackage[psamsfonts]{amssymb}
\usepackage{euscript}

\usepackage{latexsym}
\usepackage[arrow,matrix,curve]{xy}

\usepackage[hypertexnames=false]{hyperref}
\hypersetup{
    colorlinks=true,
    linkcolor=blue,
    filecolor=magenta,   
    citecolor=black,   
    urlcolor=cyan,
    }

\newskip\humongous \humongous=0pt plus 1000pt minus 1000pt

\newif\ifdtup

\def\,{\hspace{-.1cm}}
\def\hsp{,\hspace{.7cm}}

\def\ddp#1{d^2\vp_{#1}}
\def\dvx{\int d^2\vx}
\def\ddpp#1{d^2\vpp_{#1}}

\def\fc#1#2 {\frac{n}{q}#1\frac{n}{q}#2}

\def\kt{\kappa}
\def\kt{\mathfrak{K}}
\def\ks{|\kt\rangle}

\def\hp{H^{\prime(t)}}
\def\rv{{\rm{vac}}}

\newcommand{\vac}{\ensuremath{|0\rangle}}
\newcommand{\ovac}{\ensuremath{|\Omega\rangle}}
\newcommand{\ps}{\ensuremath{|\vp_0\rangle}}

\renewcommand{\sin}{\textrm{sin}}

\renewcommand{\sinh}{\textrm{sinh}}
\renewcommand{\cosh}{\textrm{cosh}}
\renewcommand{\tanh}{\textrm{tanh}}
\newcommand{\sech}{\textrm{sech}}
\newcommand{\csch}{\textrm{csch}}

\def\exp#1{\hbox{\rm exp}\left[#1\right]}

\def\sign#1{{\rm sign}\left(#1\right)}

\renewcommand{\theequation}{\arabic{section}.\arabic{equation}}
\renewcommand{\(}{\begin{equation}}
\renewcommand{\)}{end{equation} \vspace{-.05in}\linebreak}

\newcounter{saveeqn}
\newcounter{savealpheqn}

\newcommand{\alpheqn}{\setcounter{saveeqn}{\value{equation}}%
  \stepcounter{saveeqn}\setcounter{equation}{0}%
  \renewcommand{\theequation}{\mbox{\arabic{section}.\arabic{saveeqn}
\alph{equation}}}
  \renewcommand{\)}{\end{equation}}}
\def\part#1{\frac{\partial}{\partial{#1}}}%
\def\group#1{\refstepcounter{equation}\setcounter{saveeqn}
 {\value{equation}}%
  \label{#1}\setcounter{equation}{0}%
\renewcommand{\theequation}{\mbox{\arabic{section}.\arabic{saveeqn}
\alph{equation}}}
  \renewcommand{\)}{\end{equation}}}
\newcommand{\reseteqn}{\setcounter{equation}{\value{saveeqn}}%
  \renewcommand{\theequation}{\arabic{section}.\arabic{equation}}%
  \renewcommand{\)}{\end{equation}}}

\newcommand{\aalpheqn}{\setcounter{saveeqn}{\value{equation}}%
  \stepcounter{saveeqn}\setcounter{equation}{0}%
  \renewcommand{\theequation}{\mbox{
        \Alph{subsection}.\arabic{saveeqn}\alph{equation}}}
   \renewcommand{\)}{\end{equation}}}
\newcommand{\areseteqn}{\setcounter{equation}{\value{saveeqn}}%
  \renewcommand{\theequation}{\Alph{subsection}.\arabic{equation}}%
  \renewcommand{\)}{\end{equation}}}

\renewcommand{\thefootnote}{\alph{footnote}}
\renewcommand{\(}{\begin{equation}}
\renewcommand{\)}{\end{equation}}
\newcommand{\ba}{\begin{eqnarray}}
\newcommand{\ea}{\end{eqnarray}}

\renewcommand{\b}{\beta}

\renewcommand{\sl}{{\sqrt{\lambda}}}

\newcommand{\cbp}{\mathop{\vtop{\ialign{##\crcr
   $\hfil\displaystyle{}\hfil$\crcr\noalign{\kern-13pt\nointerlineskip}
   \BIG{)}\hskip0pt\crcr\noalign{\kern3pt}}}}}
\newcommand{\pa}{\mathop{\vtop{\ialign{##\crcr

$\hfil\displaystyle{\oplus}\hfil$\crcr\noalign{\kern+1pt\nointerlineskip
}
   \hspace{.08in}$^{\alpha=0}$\hskip6pt\crcr\noalign{\kern3pt}}}}}
\renewcommand{\hsp}{,\hspace{.3in}}
\newcommand{\p}{^\prime}
\newcommand{\pp}{^{\prime\prime}}

\catcode`\@=11
\def\vereq#1#2{\lower3pt\vbox{\baselineskip1.5pt \lineskip1.5pt
\ialign{$\m@th#1\hfill##\hfil$\crcr#2\crcr\sim\crcr}}}
\catcode`\@=12

\renewcommand{\(}{\begin{equation}}
\renewcommand{\)}{\end{equation}}

\def\vx{{\vec{x}}}
\def\vp{{\vec{p}}}
\def\vpp{{\vec{p}\p}}
\def\vk{{\vec{k}}}

\def\pin#1{\int \frac{d#1}{2\pi}}
\def\ppin#1{\int\hspace{-17pt}\sum \frac{d#1}{2\pi}}
\def\ppink#1{\int\hspace{-17pt}\sum\frac{d^{#1}k}{(2\pi)^{#1}}}
\def\ppinkp#1{\int\hspace{-17pt}\sum\frac{d^{#1}k\p}{(2\pi)^{#1}}}
\def\dint{\int\hspace{-12pt}\sum }
\def\pink#1{\int \frac{d^{#1}k}{(2\pi)^{#1}}}

\def\pinkp#1{\int \frac{d^{#1}k\p}{(2\pi)^{#1}}}

\def\pinvp#1{\int \frac{d^{#1}\vp}{(2\pi)^{#1}}}
\def\pinvk#1{\int\hspace{-17pt}\sum \frac{d^{#1}\vk}{(2\pi)^{#1}}}
\def\kinv#1#2{\int\hspace{-17pt}\sum \frac{d^{#1}\vec{#2}}{(2\pi)^{#1}}}
\def\pinv#1#2{\int \frac{d^{#1}\vec{#2}}{(2\pi)^{#1}}}

\def\Bd#1{B^\ddag_{k_{#1}}}
\def\Btd#1{B^{(t)\ddag}_{k_{#1}}}
\def\Bt#1{B^{(t)}_{k_{#1}}}
\def\Bdp#1{B^\ddag_{k\p_{#1}}}

\def\cc{\mathcal{C}}
\def\df{\mathcal{D}_{f}}
\def\dv{\mathcal{D}_{v}}
\def\dft{\mathcal{D}_{f}^{(t)}}
\def\dfd{\mathcal{D}_{f}^{(t)\dag}}

\def\avkd{A^\ddag_{\vec{k}}}
\def\avpd{A^\ddag_{\vec{p}}}
\def\avkm{A_{-\vec{k}}}
\def\avpm{A_{-\vec{p}}}
\def\avk{A_{\vec{k}}}
\def\avp{A_{\vec{p}}}
\def\omvk{\omega_{\vec{k}}}
\def\omvp{\omega_{\vec{p}}}

\def\navkd#1{A^\ddag_{\vec{k}_{#1}}}
\def\navpd#1{A^\ddag_{\vec{p}_{#1}}}

\def\navpm#1{A_{-\vec{p}_{#1}}}
\def\navk#1{A_{\vec{k}_{#1}}}
\def\navp#1{A_{\vec{p}_{#1}}}

\def\nomvp#1{\omega_{\vec{p}_{#1}}}

\def\I{\mathcal{I}}

\def\os{\omega_S}

\def\gx{(\gamma(x-vt))}

\def\red#1{\textcolor{red}{Jarah: #1}}
\def\blu#1{\textcolor{blue}{Baiyang: #1}}
\usepackage[dvipsnames]{xcolor}
\def\gre#1{\textcolor{ForestGreen}{Hengyuan: #1}}

\newcommand{\beas}{\begin{eqnarray*}}
\newcommand{\eeas}{\end{eqnarray*}}

\newcommand{\bquo}{\begin{quote}}
\newcommand{\enqu}{\end{quote}}

\def\lim#1{\stackrel{\rm{lim}}{{}_{#1}}}

\newcommand{\R}{{\mathbb R}}

\newcommand{\g}{\mathfrak g}

\def\ch{{\mathcal{H}}}
\def\co{{\mathcal{O}}}

\def\ok#1{\omega_{k_{#1}}}
\def\okp#1{\omega_{k\p_{#1}}}
\def\ovp#1{\omega_{\vp_{#1}}}
\def\ovpp#1{\omega_{\vpp_{#1}}}
\def\okt#1{\omega_{\kt_{#1}}}
\def\V#1{V^{(#1)}(\sqrt{\lambda}f(x))}
\def\Vg#1{V^{(#1)}(\sqrt{\lambda}f(\gamma(x-vt))}
\def\ck{\csch\left(\frac{\pi k}{2\b}\right)}

\def\mb{\mathcal{B}}
\def\mc{\mathcal{C}}
\def\md{\mathcal{D}}
\def\me{\mathcal{E}}
\def\gt{\tilde{\g}}
\def\dim{2}
\def\dimtwo{4}
\def\dimthree{6}
\def\phip{\phi}
\def\pip{\pi}

\newcommand{\beq}{\begin{equation}}
\newcommand{\eeq}{\end{equation}}
\newcommand{\bea}{\begin{eqnarray}}
\newcommand{\eea}{\end{eqnarray}}

\newskip\humongous \humongous=0pt plus 1000pt minus 1000pt

\newif\ifdtup

\catcode`\@=11

\ifthenelse{\equal{\letter}{0}}{ 

\@addtoreset{equation}{section}
\def\theequation{\arabic{section}.\arabic{equation}}

\def\@normalsize{\@setsize\normalsize{15pt}\xiipt\@xiipt
\abovedisplayskip 14pt plus3pt minus3pt%
\belowdisplayskip \abovedisplayskip
\abovedisplayshortskip \z@ plus3pt%
\belowdisplayshortskip 7pt plus3.5pt minus0pt}

\def\small{\@setsize\small{13.6pt}\xipt\@xipt
\abovedisplayskip 13pt plus3pt minus3pt%
\belowdisplayskip \abovedisplayskip
\abovedisplayshortskip \z@ plus3pt%
\belowdisplayshortskip 7pt plus3.5pt minus0pt
\def\@listi{\parsep 4.5pt plus 2pt minus 1pt
      \itemsep \parsep
      \topsep 9pt plus 3pt minus 3pt}}

\relax

\def\section{\@startsection{section}{1}{\z@}{3.5ex plus 1ex minus  .2ex}{2.3ex plus .2ex}{\large\bf}}

\def\thesection{\arabic{section}}
\def\thesubsection{\arabic{section}.\arabic{subsection}}

\def\appendix{\setcounter{section}{0}
 \def\thesection{Appendix \Alph{section}}
 \def\thesubsection{\Alph{section}.\arabic{subsection}}
 \def\theequation{\Alph{section}.\arabic{equation}}}
\renewcommand{\theequation}{\arabic{section}.\arabic{equation}}

}{
\renewcommand{\theequation}{\arabic{equation}}

} 

\begin{document}
\def\thefootnote{\fnsymbol{footnote}}
\def\thetitle{A Finite Tension for the $\phi^4_4$ Domain Wall}
\def\autone{Jarah Evslin}
\def\autthree{Hui Liu}
\def\autfour{Baiyang Zhang}
\def\auttwo{Hengyuan Guo}
\def\affd{Yerevan Physics Institute, 2 Alikhanyan Brothers St., Yerevan, 0036, Armenia}
\def\affb{University of the Chinese Academy of Sciences, YuQuanLu 19A, Beijing 100049, China}
\def\affa{Institute of Modern Physics, NanChangLu 509, Lanzhou 730000, China}
\def\affc{School of Physics and Astronomy, Sun Yat-sen University, Zhuhai 519082, China}
\def\affe{Institute of Contemporary Mathematics, School of Mathematics and Statistics,Henan University, Kaifeng, Henan 475004, P. R. China}


\ifthenelse{\equal{\pr}{1}}{
\title{\thetitle}
\author{\autone}
\author{\auttwo}
\author{\autthree}
\affiliation {\affa}
\affiliation {\affb}
\affiliation {\affc}

}{

\begin{center}
{\large {\bf \thetitle}}

\bigskip

\bigskip


{\large 
\noindent  \autone{${}^{12}$}
\footnote{jarah@impcas.ac.cn},
\auttwo{${}^{3}$}
\footnote{guohy57@mail.sysu.edu.cn }, 
\autthree{${}^{4}$} \footnote{hui.liu@yerphi.am}, 
\autfour{${}^{5}$\footnote{byzhang@henu.edu.cn}}
}


\vskip.7cm

1) \affa\\
2) \affb\\
3) \affc\\
4) \affd\\
5) \affe\\

\end{center}

}

\begin{abstract}
\noindent

\noindent
In 1+1 dimensions, it is well known that the quantum states corresponding to solitons are well described by coherent states.  In his 1975 Erice lectures, Coleman observed that this construction does not extend to higher dimensions, as the coherent states have infinite energy density.  He challenged the students to construct the quantum states corresponding to solitons in higher dimensions, a problem which remains unsolved today.  However, 
even in 1+1 dimensions, the correct quantum states are actually given by deformations of coherent states.  In the 3+1 dimensional $\phi^4$ double-well model, we show that the leading deformation, which is just a squeeze, already cancels the one-loop divergence in the energy density of the domain wall soliton.  
\end{abstract}
%
\setcounter{footnote}{0}
\renewcommand{\thefootnote}{\arabic{footnote}}

\ifthenelse{\equal{\pr}{1}}
{
\maketitle
}{}

\section{Introduction}

If a classical field theory exhibits a soliton solution, then generically the Hilbert space of the quantum field theory will exhibit a solitonic sector.  In other words, in addition to the usual Fock space of perturbative excitations of the vacuum, there will be an additional Fock space of perturbative excitations of the soliton.  Ultraviolet divergences appear in both the vacuum and soliton sectors.  However, both are described by the same Hamiltonian which has the same set of counterterms.  Thus the renormalization conditions may be applied only once, and so a soliton sector only exists if the same counterterms eliminate ultraviolet divergences in both sectors.

There are many convincing arguments in the literature that this indeed occurs.  It was noted at one loop in 1+1 dimensions in Ref.~\cite{gjs75}, but postulated to hold for all renormalizable theories in Ref~\cite{polyakov75}.  A diagramatic argument that divergences in the S-matrix are eliminated was presented in Ref.~\cite{fk77}.  However, Coleman has argued \cite{erice} that this is not the case if one constructs the soliton sector using the usual coherent state construction of Refs.~\cite{vinc72,cornwall74,taylor78}.  This conventional construction in fact leads to an infinite energy density for states in the soliton sector for field theories in more than 1+1 dimensions.  This leads one to ask just how the solitonic sector is to be constructed beyond 1+1 dimensions for this divergence to be avoided.

As has been recently emphasized in Ref.~\cite{cocorr23}, the coherent state corresponding to a soliton is somewhat deformed.  The leading deformation is a squeeze that puts all bound and continuum normal modes in their ground state and averages over zero modes.  In the present note, we show that this squeeze is already sufficient to eliminate the simplest manifestation of the divergence discussed by Coleman, that appearing at one loop in the domain wall soliton \cite{okun} of the 3+1 dimensional $\phi^4$ double-well model.

\section{The Domain Wall Soliton Tension}

Consider the 3+1 dimensional theory of a scalar $\phi(\vx)$ and its conjugate momentum $\pi(\vx)$ described by the Hamiltonian
\beq
H=\int d^3\vx:\ch(\vx):\hsp
\ch(\vx)=\frac{\pi^2(\vx)+\sum_{i=1}^3\left(\partial_i \phi(\vx)\right)^2}{2}+\frac{\lambda_0 \phi^4(\vx)-m_0^2 \phi^2(\vx)}{4}+A.
\eeq
Here the normal ordering $::$ is defined at the mass scale $m_0$, and the $c$-number $A$ is a counterterm.  

Classically, there are two minimum energy configurations given by $\phi(\vx)=\pm m_0/\sqrt{2\lambda_0}$.  There is also a domain wall soliton solution 
\beq
\phi(\vx)=\frac{m_0}{\sqrt{2\lambda_0}}{\rm{tanh}}\left(\frac{m_0 x_1}{2}\right)
\eeq
with tension
\beq
\rho_0=\frac{m_0^3}{3\lambda_0}.
\eeq
The profile in the $x_1$ direction is just that of the $\phi^4$ kink, whose mass is $\rho_0$, while it is independent of the other directions.  Such higher-dimensional extensions of the kink have been of interest lately, even finding applications in cosmology \cite{bp21,bp23}.

A simple formula for the one-loop correction to the kink mass was provided in Ref.~\cite{cahill76}.   It was shown in Ref.~\cite{me3d} that essentially this same formula provides the one-loop correction to the tension of the domain wall soliton
\beq
\rho_1=-\frac{1}{4}\ppin{k_1}\pinv{3}{p} |\gt_{-k_1}(p_1)|^2 \frac{(\omega_{k_1p_2p_3}-\omega_{p_1p_2p_3})^2}{\omega_{p_1p_2p_3}}\hsp \omega_{pqr}=\sqrt{m^2+p^2+q^2+r^2}. \label{c76}
\eeq
This derivation explicitly used the construction of the kink as a deformed, coherent state.

Here $\dint$ sums over the bound normal modes of the kink and integrates over the continuum modes.  $\gt_k(p)$ is the Fourier transform of the normal mode indexed by $k$.

More precisely, Ref.~\cite{me3d} considered the two-dimensional case, where this expression is finite.  In fact, the tension of the 2+1 dimensional domain wall was found explicitly in Ref.~\cite{zar98}.  However, in 3+1 dimensions, $\rho_1$ enjoys an ultraviolet divergence arising from the continuum normal modes
\beq
\tilde{\g}_k(p)=\frac{2k^2-m_0^2}{\ok{}\sqrt{m_0^2+4k^2}}2\pi\delta(p+k)+\frac{6\pi p}{\ok{}\sqrt{m_0^2+4k^2}} \csch\left(\frac{\pi (p+k)}{m_0}\right).
\eeq
The Dirac $\delta$ term does not contribute to the tension, as it is multiplied by a double zero.  However the other term leads to a contribution to $\rho_1$ of 
\beq
-\frac{3\pi}{2}\pin{k_1}\frac{\omega_{k_1}}{m_0^2+4k_1^2}\pin{p_1}p_1^2\csch^2\left[\frac{\pi(p_1-k_1)}{m_0}\right]\left( 2-3\frac{\omega_{p_1}}{\ok1}+\frac{\omega^3_{p_1}}{\ok{1}^3}\right)
\eeq
where we ignore the discrete modes $k_1$, which do not lead to ultraviolet divergences.
At large $|k_1|$ this contribution to $\rho_1$ tends to
\beq
-\frac{3m_0^3}{16\pi}\pin{k_1} \frac{1}{|k_1|}.
\eeq
An ultraviolet cutoff $|k_1|\leq \Lambda$ then leads to a leading divergence of
\beq
\rho_1=-\frac{3m_0^3}{16\pi^2}{\rm{ln}}(\Lambda).
\eeq
The total tension is $\rho_0+\rho_1$.  Is this divergent at $\Lambda\rightarrow\infty$?  Not necessarily, as $m_0$ and $\lambda_0$ also diverge.  To answer this question, let us expand in powers of $\lambda$
\beq
m_0^2=m^2-\delta m^2+O(\lambda^2)\hsp
\sqrt{\lambda_0}=\sqrt{\lambda}-\delta\sqrt\lambda+O(\lambda^{5/2})
\eeq
where $\delta m^2$ and $\delta\sqrt\lambda$ are of order $O(\lambda)$ and $O(\lambda^{3/2})$ respectively.  At this stage we have not fixed a renormalization condition, however we assume that a renormalization condition is chosen such that $m$ and $\lambda$ are finite.

Then, up to order $O(\lambda^0)$, dropping finite terms, the tension is
\beq
\rho_0+\rho_1=\frac{(m^2-\delta m^2)^{3/2}}{3(\sqrt{\lambda}-\delta\sqrt{\lambda})^2}-\frac{3m^3}{16\pi^2}{\rm{ln}}(\Lambda)=\frac{m^3}{3\lambda}-\frac{m\delta m^2}{2\lambda}+\frac{2m^3\delta\sqrt\lambda}{3\lambda^{3/2}}-\frac{3m^3}{16\pi^2}{\rm{ln}}(\Lambda). \label{ten}
\eeq
The first term in the rightmost expression is of order $O(1/\lambda)$ and is manifestly finite.  The question is therefore whether the sum of the last three terms is finite.

\section{Renormalizing a Vacuum Sector}
We have now reduced the problem of finding the domain wall tension to the problem of finding the counterterms $\delta m^2$ and $\delta\sqrt\lambda$.  These counterterms must not only eliminate divergences, for example in $\rho_0+\rho_1$, in the soliton sector, but also in the vacuum sector.  Therefore they may be fixed uniquely through a calculation in the vacuum sector.

Needless to say, the renormalization of the $\phi^4$ theory in the vacuum sector is an old and well-understood subject  \cite{kleinert}.  There are many different renormalization prescriptions that lead to answers whose finite parts disagree but whose infinite parts agree.  Furthermore, it has been appreciated for over 50 years that the spontaneous symmetry breaking of this theory does not affect the infinite parts of the counterterms.  We will therefore not dwell on this problem, but simply review the main steps.

First of all, at each order, $A$ is chosen to make the vacuum energy finite.  With this choice, contributions arising from diagrams with disconnected components with no external legs will be, up to finite terms, canceled by contributions from $A$.

\begin{figure}[htbp]
\centering
\includegraphics[width = 0.75\textwidth]{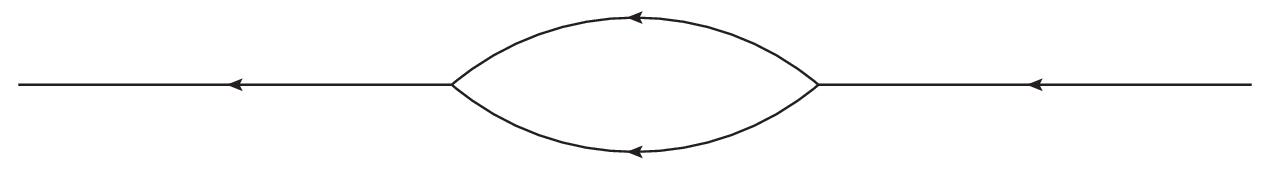}
\caption{This divergent diagram contributes to $\delta m^2/m^2-6\delta\sqrt\lambda/\sqrt\lambda$}\label{m2fig}
\end{figure}

Next, the divergence in the propagator can be fixed using a renormalization condition.  This invariably involves the ultraviolet divergent diagram in Fig.~\ref{m2fig} which, dropping finite terms, fixes
\beq
\frac{\delta m^2}{m^2}-\frac{6\delta\sqrt\lambda}{\sqrt\lambda}=\frac{9\lambda}{2}\int\frac{d^3\vp}{(2\pi)^3}\frac{1}{|\vp|^3} \label{primeq}.
\eeq
To see this, note that after shifting $\phi$ by the vacuum expectation value $v=-m/
\sqrt{2\lambda}$, the $\phi^2$ term in the Hamiltonian density is equal to the usual $m^2\phi^2/2$ plus a counterterm.
\beq
\left(\frac{\delta m^2}{4}-\frac{3m^2\delta\sl}{2\sl}\right)\phi^2.
\eeq

This coefficient is proportional to the left hand side of (\ref{primeq}) and it needs to eliminate the divergence in Fig.~\ref{m2fig}, which appears on the right hand side of Eq.~(\ref{primeq}).
The same ultraviolet cutoff yields
\beq
\frac{\delta m^2}{m^2}-\frac{6\delta\sqrt\lambda}{\sqrt\lambda}=\frac{9\lambda}{4\pi^2}\rm{ln}(\Lambda)+\rm{finite}. \label{m2eq}
\eeq

\begin{figure}[htbp]
\centering
\includegraphics[width = 0.75\textwidth]{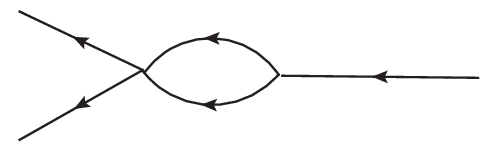}
\caption{This divergent diagram contributes to $\delta\sqrt\lambda$}\label{dlfig}
\end{figure}

One next uses the third and final renormalization condition 
to fix the three point interaction which at one loop contains the divergent contribution in Fig.~\ref{dlfig} as well as reducible divergences that are canceled by (\ref{m2eq}) and disconnected contributions that are canceled by $A$.  Again applying an ultraviolet cutoff, this diagram leads to
\beq
{\delta\sqrt\lambda}=-\frac{9\lambda^{3/2}}{8}\int\frac{d^3\vp}{(2\pi)^3}\frac{1}{|\vp|^3}+\rm{finite}=-\frac{9\lambda^{3/2}}{16\pi^2}\rm{ln}(\Lambda)+\rm{finite}.
\eeq
Inserting this into (\ref{m2eq}) one finds
\beq
\delta m^2=-\frac{9m^2\lambda}{8\pi^2}\rm{ln}(\Lambda)+\rm{finite}.
\eeq

Inserting this and (\ref{m2eq}) into (\ref{ten}), one sees that the divergent parts of the three last terms indeed cancel, and so the domain wall tension $\rho_0+\rho_1$ is finite at one loop.

\section{Remarks}

Recall that Ref.~\cite{me3d} derived the one-loop tension (\ref{c76}) from the construction $\df\vac_0$ where $\df$ was the displacement operator that created a coherent state.  Here $\vac_0$ was the tree level vacuum, squeezed so that, instead of being annihilated by the operators that destroy plane waves, it is annihilated by the operators that destroy normal modes except for the zero mode.  In the case of the zero mode, it was annihilated by the momentum operator for the center of mass.  In other words, it is a Bogoliubov transformation of the true vacuum $|\Omega\rangle_0$ \cite{wentzel}, and so is described by a squeeze, which was explicitly constructed in Ref.~\cite{mestate}.

Therefore we may summarize our results as follows.  Coleman's statement that the coherent state $\df|\Omega\rangle_0$ has infinite energy density is true.  The corresponding tension is $\rho_0$, which indeed is infinite.  However the correct ground state is, at $O(\lambda^0)$, given by $\df\vac_0$ which is a squeezed coherent state.  This state has finite energy density and leads to a finite tension for the domain wall soliton.  This resolves the puzzle raised by Coleman at one loop, in a scalar theory.  It remains to test it at higher loops and with more general matter content.

The full quantum $\phi^4$ theory is believed to be trivial, in the sense that field renormalization makes the theory free as $\Lambda\rightarrow\infty$ for any fixed $m_0$ and $\lambda_0$ \cite{froh82,ai21}.  However, we consider the theory at one loop, where field renormalization is not necessary.  Alternately, our results may be stated as follows.  The domain wall tension may be written as a function of $m,\ \lambda$\ and $\Lambda$.  We find, at one loop, that as $\Lambda$ increases with $m$ and $\lambda$ fixed, the tension is bounded.  This is true despite the fact that, as $\Lambda$ increases, the bare parameters are unbounded.

\section* {Acknowledgement}

\noindent
JE is supported by NSFC MianShang grants 11875296 and 11675223.  JE and HL are supported by the Ministry of Education, Science, Culture and Sport of the Republic of Armenia under the Remote Laboratory Program, grant number 24RL-1C047.

\end{document}